\documentclass[a4paper, 11pt]{article}

\usepackage{graphicx}
\usepackage{amsmath}
\usepackage{amsfonts}
\usepackage{amssymb}
\usepackage{authblk}

\usepackage{amsthm,amsmath,natbib}
\RequirePackage{hyperref}
\usepackage{multirow}
\usepackage{amsmath,amssymb}
\DeclareMathOperator*{\plim}{plim}
\newtheorem{Ex}{Example}
\usepackage{pdfpages}
\usepackage{caption}
\numberwithin{equation}{section}
\theoremstyle{plain}

\emergencystretch=\maxdimen
\usepackage{tikz}
\usetikzlibrary{arrows,backgrounds,snakes,patterns}
\usetikzlibrary{shapes,arrows,chains}
\usepackage{verbatim}
\usepackage{booktabs}

\usepackage[flushleft]{threeparttable}

\begin{document}
\pdfoutput=1
\thispagestyle{empty} 

\title{Model choice in separate families: A comparison between the FBST and the Cox test}

\author[1]{Cachimo Combo Assane \footnote{cachimo.assane@gmail.com}}
\author[1]{Basilio de Bragan\c{c}a Pereira \footnote{basilio@hucff.ufrj.br}}
\author[2]{Carlos Alberto de Bragan\c{c}a Pereira \footnote{cpereira@ime.usp.br}}

\affil[1]{Universidade Federal do Rio de Janeiro (UFRJ), Rio de Janeiro, Brazil}
\affil[2]{Universidade de S\~{a}o Paulo (USP), S\~{a}o Paulo, Brazil}

\maketitle

\begin{abstract}
Tests of separate families of hypotheses were initially considered by \cite{Cox61,Cox62}. In this work, the Fully Bayesian Significance Test, FBST, is evaluated for discriminating between the lognormal, gamma and Weibull models whose families of distributions are separate. Considering a linear mixture model including all candidate distributions, the FBST tests the hypotheses on the mixture weights in order to calculate the evidence measure in favor of each one. Additionally, the density functions of the mixture components are reparametrized in terms of the common parameters, mean $\mu$ and variance $\sigma^2$ of the population, since the comparison between the models is based on the same dataset, i.e, on the same population. Reparametrizing the models in terms of the common parameters also allows one to reduce the number of the parameters to be estimated. In order to evaluate the performance of the procedure, some numerical results based on simulated sample points are given. In these simulations, the results of FBST are compared with those of the Cox test. Two applications examples illustrating the procedure for uncensored dataset are also presented.\\ \\

\end{abstract}
\section{Introduction}\label{sec1}

An important problem in statistical analysis is that of the choice between alternative statistical models. The Neyman-Pearson theory of hypothesis testing applies only if the models belong to the same family of distributions. Alternatively, special procedures are required if the models belong to families that are separate (or nonnested) in the sense that an arbitrary member of one family cannot be obtained as a limit of members of the other. The set of separate families of probability distributions includes the lognormal, gamma and Weibull models, which play an important roles in survival and reliability analysis \citep{Pereira81,Lawless02}.

A considerable amount of research on separate families of hypotheses has been developed since the fundamental work of \citet{Cox61, Cox62}, who first dealt with the problem. For reviews and references, see \citet{Pereira05, Araujo05, Araujo07,Pereira17}.

Significance tests are regarded as procedures for measuring the consistency of data with a null hypothesis \citep{Cox77,Kempt76}. \citet{Berger87} consider the classical p-value as a measure of evidence of the null hypothesis and present alternative Bayesian measures of evidence, the Bayes Factor and the posterior probability of the null hypothesis.  However, as discussed in \citet{SternP14}, p-values are a tail area under the null hypothesis, calculated in the sample space, not in the parameter space where the hypothesis is formulated. \cite{kamary14} has documented some difficulties with traditional Bayesian tests and Bayesian model choices via posterior probabilities. The Bayesian analysis also encounters difficulties in using Bayes factors \citep{Araujo05,Araujo07}. For example, when prior information is weak and improper prior is applied, then, the usual Bayes factor is not well defined. To overcome these difficulties duo to improper priors, modified Bayes factors have been proposed \citep{Araujo07,Pereira17}.

The Fully Bayesian Significance Test (FBST) is presented by \citet{PereiraS} as alternative to the Bayes factor and classical p-values for precise hypotheses. The basis for the FBST is an index known as the e-value (where e stands for evidence), which measures the inconsistency of the hypothesis  using several parameter points together with the posterior densities. For reviews and further references on FBST, see \cite{PereiraSW} and \citet{SternP14}. For a few interesting applications illustrating the use of e-values and the FBST to practical problems, see \citet{Diniz12}, \citet{Lauretto03}, \citet{Lauretto05}, \citet{Lauretto09} and \citet{PereiraS}.

In this paper we consider the FBST for discriminating between the lognormal, gamma and Weibull models whose families of distributions are separate. As suggested by \citet{Cox61}, we  analyze this problem in the context of linear mixture of the candidate models. That is, the models under discrimination are considered as components of a linear mixture model. The FBST procedure is used for testing the hypotheses on the mixture weights in order to calculate the evidence measure in favor of each model.

The novelty of our work is that the density functions of the mixture components are reparametrized in terms of the mean $\mu$ and the variance $\sigma^2$ of the population so that the models under discrimination share common parameters \citep{kamary14,Pereira17}. A standard Bayesian approach to finite mixture models is to consider different pairs of parameters for each of these models and to adopt independent prior distributions for each pair of parameters and a Dirichlet prior on the mixture weights \citep{Lauretto05,Lauretto07}. However, since the comparison between the models is based on the same dataset, i.e, on the same population, we believe that it would be inappropriate to consider different means and variances for these models and, in addition, with different prior distributions. An additional advantage of this reparametrization is that it allows one to reduce the number of the parameters to be estimated.

To illustrate the procedure, numerical results based on simulations of sample points were presented. In these simulations, empirical results for comparison between FBST and Cox test are discussed. We also applied the lognormal-gamma-Weibull mixture model to the simulated data in order to evaluate the performance of the FBST on identifying the true distribution of the generated sample. Two application examples illustrating the procedure for uncensored dataset are also presented.

The plan of the paper is as follows. Section \ref{sec2} presents a brief review of Cox test. Section \ref{sec3} reviews the basic concepts o FBST. Section \ref{sec4} discusses the FBST formulation for discriminating between separate models in the context of mixture models. Section \ref{sec5} presents the simulation results of both FBST and Cox test. In Section \ref{sec6} two real datasets are used as examples to illustrate the procedures. Final remarks are presented in Section \ref{sec7}.

\section{The Cox test} \label{sec2}
Let $y=(y_1, ..., y_n)$ be independent and identically distributed observations from some unknown distribution $F$. Suppose that there are null hypothesis, $H_f:F\in \mathfrak{F}_f$, where $\mathfrak{F}_f$ is a family of probability distributions with density $f(y|\alpha)$ and alternative hypothesis, $H_g: F\in\mathfrak{F}_g$, where $\mathfrak{F}_g$ is another family of probability distributions with density $g(y|\beta)$. Hence $\alpha$ and $\beta$ are unknown parameter vectors and it is assumed further that the families of $f$ and $g$ are separate in sense defined above. Formal definitions of separate hypotheses are given in \citet{Pereira17}.

The asymptotic test developed by \citet{Cox61,Cox62} is based on a modification of the Neyman--Pearson maximum likelihood ratio. The test statistic for $H_f$ against $H_g$ is

\begin{equation*}\label{Coxtest}
T_{fg}=\ell_f(\hat{\alpha})-\ell_g(\hat{\beta})-n\left[\displaystyle\plim_{n\to\infty}\frac{\ell_f(\hat{\alpha})-\ell_g(\hat{\beta})}{n}\right]_{\alpha=\hat{\alpha}},
\end{equation*}
where $\ell_f(\hat{\alpha})$ and $\ell_g(\hat{\beta})$ are the maximized log-likelihoods under $H_f$ and $H_g$, respectively; $\hat{\alpha}$ and $\hat{\beta}$ denote the maximum likelihood estimates; plim\index{plim} represents convergence in probability; and the subscript $\alpha$ indicates that the means are calculated under $H_f$.

Cox showed that, asymptotically, under the alternative hypothesis, $T_{fg}$ has a negative mean,  whereas, under the null hypothesis, it is normally distributed with mean zero and variance
\begin{equation*}\label{CoxVar}
V_{\alpha}(T_{fg})=V_{\alpha}\left\{\ell_f(\alpha)-\ell_g(\beta_{\alpha})\right\}-C_{\alpha}'I_{\alpha}^{-1}C_{\alpha},
\end{equation*}
where $\beta_\alpha$ is the probability limit of $\hat{\beta}$ under $H_f$, as $n\longrightarrow\infty$, $C_\alpha\equiv n\frac{\partial}{\partial\alpha}\left[\displaystyle\plim_{n\to\infty}\frac{\ell_f(\hat{\alpha})-\ell_g(\hat{\beta})}{n}\right]$, and $I_{\alpha}$ the information matrix of $\alpha$. When $H_g$ is the null hypothesis and $H_f$ is the alternative hypothesis, analogous results are obtained for a statistic $T_{gf}$. Therefore, $T_{fg}^{*}=T_{fg}\left\{V(T_{fg})\right\}^{-1/2}$ and $T_{gf}^{*}=T_{gf}\left\{V(T_{gf})\right\}^{-1/2}$ under $H_f$ and $H_g$, respectively, are approximately standard normal variables, and two-tailed tests can be performed. The possible outcomes when both tests are performed are presented in \citet{Pereira17}.

As an illustration of the calculations for the Cox's test statistics, following \citet{Pereira78}, suppose that $H_f$ specifies that the distribution is lognormal and $H_g$ specifies that it is Weibull; that is

\begin{align*}
H_f&: f(y|\alpha)=\frac{1}{y\sqrt{2\pi\alpha_2}}\exp\left\{-\frac{(\log y-\alpha_1)^2}{2\alpha_2}\right\}, \ \alpha=(\alpha_1, \alpha_2),& \\\vspace*{1cm}
H_g&: g(y|\beta)=\frac{\beta_2}{\beta_{1}^{\beta_2}}y^{\beta_2-1}\exp\left\{-\left(\frac{y}{\beta_1}\right)^{\beta_2}\right\}, \ \beta=(\beta_1,\beta_2).&
\end{align*}

We then have

\begin{align*}
T_{fg}&=n\left\{\hat{\beta_2}\ln\hat{\beta}_1-\beta_{2\hat{\alpha}}\ln\beta_{1\hat{\alpha}}-\ln\hat{\beta_2}+\ln\beta_{2\hat{\alpha}}-\hat{\alpha}_1(\hat{\beta}_2-\beta_{2\hat{\alpha}})\right\} \ \ \mbox{and}&\\\vspace*{1cm}
V_\alpha(T_{fg})&=0.2183n,&
\end{align*}
where $\beta_{1\hat{\alpha}}=\exp\{\hat{\alpha}_1+\sqrt{\hat{\alpha}_2}/2\}$ and $\beta_{2\hat{\alpha}}=\hat{\alpha}_2^{-1/2}$ are the estimated values of $\beta_{1\alpha}$ and $\beta_{2\alpha}$ which are the probability limits of $\hat{\beta}_1$ and $\hat{\beta}_2$ under $H_f$, respectively.

Also

\begin{align*}
T_{gf}&=n\left\{\hat{\beta}_2(\hat{\alpha}_1-\alpha_{1\hat{\beta}})+\frac{1}{2}\ln\frac{\hat{\alpha}_2}{\alpha_{2\hat{\beta}}}\right\} \ \ \mbox{and}& \\ \vspace*{0.8 cm}
V_\beta(T_{gf})&=0.2834n,&
\end{align*}
where $\alpha_{1\hat{\beta}}=-0.5772/\hat{\beta}_2+\ln\hat{\beta}_1$ and $\alpha_{2\hat{\beta}}=1.6449/\hat{\beta}_{2}^{2}$ are the estimated values of $\alpha_{1\beta}$ and $\alpha_{2\beta}$, which are the probability limits of $\hat{\alpha}_1$ and $\hat{\alpha}_2$ under $H_g$, respectively.

The Cox's test statistics for discriminating between exponencial $vs.$ lognormal, lognormal $vs.$ gamma and gamma $vs.$ Weibull distributions can be found in \citet{Pereira17}.

\section{Fully Bayesian Significance Test (FBST)}\label{sec3}

The FBST of \citet{PereiraS}, which is reviewed in \citet{PereiraSW}, is a Bayesian version of significance testing, as considered by \citet{Cox77} and \citet{Kempt76}, for precise (or sharp) hypotheses.

First, let us consider a real parameter $\theta$, a point in the parameter space $\Theta\subset\Re$, and an observation $y$ of the random variable $Y$. A frequentist looks for the set $I\in\Re$ of sample points that are at least as inconsistent with the hypothesis as $y$ is. A Bayesian looks for the tangential set $T(y)\subset\Theta$ \citep{PereiraSW}, which is a set of parameter points that are more consistent with the observed $y$ than the hypothesis is. An example of a sharp hypothesis in a parameter space of the real line is of the type $H: \theta=\theta_0$. The evidence value in favor of $H$ for a frequentist is the usual p-value, $P(Y\in I|\theta_0)$, whereas for a Bayesian, the evidence in favor of $H$ is the e-value, $ev=1-\mbox{Pr}(\theta\in T(y)|y)$.

In the general case of multiple parameters, $\Theta\subset\Re^k$, let the posterior distribution for $\theta$ given $y$ be denoted by $q(\theta|y)\propto \pi(\theta)L(y,\theta)$, where $\pi(\theta)$ is the prior probability density of $\theta$ and $L(y,\theta)$ is the likelihood function. In this case, a sharp hypothesis is of the type $H:\theta\in\Theta_H\subset\Theta$, where $\Theta_H$ is a subspace of smaller dimension than $\Theta$. Letting $\displaystyle\sup_H$ denote the supremum of $\Theta_H$, we define the general Bayesian evidence and the tangential set, $T(y)$, as follows:
\begin{equation}\label{tangent}
q^{*}=\displaystyle\sup_Hq(\theta|y) \ \ \mbox{and} \ \ T(y)= \{\theta: q(\theta|y)>q^{*} \}.
\end{equation}
The Bayesian  evidence value against $H$ is the posterior probability of $T(y)$,
\begin{equation}\label{ev}
\overline{ev}=\mbox{Pr}(\theta\in T(y)|y)=\int_{T(y)}q(\theta|y)d\theta; \ \ \mbox{consequently}, \ \ ev=1-\overline{ev}.
\end{equation}

It is important to note that evidence that favors $H$ is not evidence against the alternative, $\overline{H}=\Theta\setminus H$, because it is not a sharp hypothesis. This interpretation also holds for p-values in the frequentist paradigm. As in \citet{PereiraSW}, we would like to point out that this Bayesian significance index uses only the posterior distribution, with no need for additional artifacts such as the inclusion of positive prior probabilities for the hypotheses or the elimination of nuisance parameters. The computation of the e-values does not require asymptotic methods, and the only technical tools needed are numerical optimization and integration methods.

Let us consider the distribution function of the evidence value against the hypothesis, $\overline{V}(c)=\mbox{Pr}(\overline{\mbox{ev}}\leq c)$, given $\theta^0$, the true value of the parameter. Under appropriate regularity conditions, for increasing sample size, $n\rightarrow\infty$, we can state the following:
\begin{itemize}
  \item If $H$ is false, $\theta^0\not\in H$, then $\overline{ev}$ converges (in probability) to $1$, that is, $\overline{V}(0<c<1)\rightarrow0$.
  \item if $H$ is true, $\theta^0\in H$, then $\overline{V}(c)$, the confidence level, is approximated by the function

\begin{equation*}
Q(t, h, c)=F_{t-h}[F_{t}^{-1}(c)],
\end{equation*}
where $t=\mbox{dim}(\Theta)$, $h=\mbox{dim}(H)$, $F_g(x)$ is the cumulative density function of chi-square distribution with $g$ degree of freedom.
\end{itemize}

Hence, for large $n$, to reject $H$ with level of significance $\gamma$, we set $c$ such that $Q(t, h, c)=1-\gamma$, i.e., $c=F_{t}[F_{t-h}^{-1}(1-\gamma)]$. Therefore, the FBST procedure rejects $H$ if $\overline{\mbox{ev}}(H)>c$.

\citet{Diniz12} have shown that, asymptotically, there is a relationship between  $\overline{\mbox{ev}}(H)$ and p-value based on the Likelihood ratio test. Thus, from asymptotic normality property, $\overline{\mbox{ev}}(H)\approx F_{t}[F_{t-h}^{-1}(1-\mbox{p-value})]$. We then have
\begin{equation}
\mbox{p-value}=1-F_{t-h}[F_{t}^{-1}(\overline{\mbox{ev}}(H))].
\end{equation}
\section{Mixture of separate models}\label{sec4}

Let us consider a dataset $y=\{y_1, \ldots, y_n\}$ and $m$ alternative probability distributions with densities $f_1(y|\psi_1),f_2(y|\psi_2),\ldots, f_m(y|\psi_m)$. Here, $\psi_k, k=1,\ldots,m$, are unknown (vector) parameters and the families of distributions are separate. The problem of interest is to measure the evidence in favor of each model for fitting the dataset. As suggested by \citet{Cox61}, we can consider a general model including all candidate distributions where the choice of a specific distribution is a special case. In this work, we formulate the FBST for the linear mixture of separate models as a selection procedure. Denoting $\boldsymbol{\theta}=(\psi_1,\ldots,\psi_m, \boldsymbol{p})$, the density function for $m-$component mixture model is
\begin{equation} \label{mixture model}
f(y_j|\boldsymbol{\theta})=p_1f_1(y_j|\psi_1)+\ldots+p_mf_m(y_j|\psi_m) \ \ p_k\geq0, \ \displaystyle\sum_{k=1}^{m}p_k=1.
\end{equation}
where $\boldsymbol{p}=(p_1,\ldots,p_m)$ is the vector of the mixture weights.

In this paper, the density functions of the mixture components in (\ref{mixture model}) are reparametrized in terms of the mean $\mu$ and the variance $\sigma^2$ of the population so that the models under comparison share common parameters \citep{kamary14,Pereira17}. The main reason for this reparametrization is that, since the comparison between the models is based on the same dataset, i.e, on the same population, we believe that it would be inappropriate to consider different means and variances for these models and, in addition, with different prior distributions as is commonly performed in traditional Bayesian approach to finite mixture model. Therefore, we have $\boldsymbol{\theta}=(\mu,\sigma^2, \boldsymbol{p})$ denoting all parameters of the mixture model, where $\mu$ and $\sigma^2$ are the connecting parameters, with $\boldsymbol{p}$ corresponding to the vector of the mixture weight.

Assuming that the $y_i$ are conditionally (on the parameter) independent and identically distributed, then, the likelihood function is
\begin{equation} \label{verossP}
L(y,\boldsymbol{\theta})=\displaystyle\prod_{j=1}^{n}\sum_{k=1}^{m}p_kf_k(y_j|\mu,\sigma).
\end{equation}

The families of distributions considered include the lognormal, gamma and Weibull models. Hence, the relationship between the parameters of these models through the $\mu$ and $\sigma^2$ is described as follows.

\begin{itemize}
  \item [(i)] Let $y$ be a $\mbox{lognormal}(\alpha_1,\alpha_2), \alpha_1\in \ \mathbb{R} \ \mbox{and} \ \alpha_2>0$, with probability density function
  \begin{equation*}
  f_L(y|\alpha_1,\alpha_2)=\frac{1}{y\sqrt{2\pi\alpha_2}}\exp\left\{-\frac{(\log y-\alpha_1)^2}{2\alpha_2}\right\}.
  \end{equation*}
 We then have
\begin{equation}\label{rep_lnormal}
\left\{
  \begin{array}{ll}
 \vspace*{0.3 cm}
    \mu= E(y|\alpha_1,\alpha_2)=\mbox{e}^{\alpha_1+\alpha_2/2}\\
    \sigma^2=Var(y|\alpha_1,\alpha_2)=(\mbox{e}^{\alpha_2}-1)\mbox{e}^{2\alpha_1+\alpha_2}
  \end{array}
\right. \Rightarrow\left\{
                     \begin{array}{ll}
                         \vspace*{0.3 cm}
                       \alpha_1=\log\frac{\mu^2}{\sqrt{\mu^2+\sigma^2}}\\
                       \alpha_2=\sqrt{\log\frac{\mu^2+\sigma^2}{\mu^2}}.
                     \end{array}
                   \right.
\end{equation}

  \item [(ii)] Let $y$ be a $\mbox{gamma}(\gamma_1,\gamma_2), \gamma_1>0 \ \mbox{and} \ \gamma_2>0$, with probability density function
  \begin{equation*}
  f_G(y| \gamma_1,\gamma_2)=\frac{1}{\Gamma(\gamma_2)\gamma_1^{\gamma_2}}y^{\gamma_2-1}\exp\left\{-\frac{y}{\gamma_1}\right\}.
  \end{equation*}
Therefore
\begin{equation}\label{rep_gama}
\left\{
  \begin{array}{ll}
 \vspace*{0.3 cm}
    \mu= E(y|\gamma_1,\gamma_2)=\gamma_1\gamma_2\\
    \sigma^2=Var(y|\gamma_1,\gamma_2)=\gamma_2\gamma_{1}^{2}
  \end{array}
\right. \Rightarrow\left\{
                     \begin{array}{ll}
                         \vspace*{0.3 cm}
                       \gamma_1=\frac{\sigma^2}{\mu}\\
                       \gamma_2=\frac{\mu^2}{\sigma^2}.
                     \end{array}
                   \right.
\end{equation}

  \item [(iii)] When $y\sim\mbox{Weibull}(\beta_1,\beta_2),\beta_1>0 \ \mbox{and} \ \beta_2>0$, with probability density function
\begin{equation*}
 f_W(y|\beta_1,\beta_2)=\frac{\beta_2}{\beta_{1}^{\beta_2}}y^{\beta_2-1}\exp\left\{-\left(\frac{y}{\beta_1}\right)^{\beta_2}\right\},
 \end{equation*}
then
\begin{align}\label{rep_weibull}
&\left\{
  \begin{array}{ll}
 \vspace*{0.3 cm}
    \mu= E(y|\beta_1,\beta_2)=\beta_1\Gamma(1+1/\beta_2)\\
    \sigma^2=Var(y|\beta_1,\beta_2)=\beta_{1}^{2}\Gamma(1+2/\beta_2)-\beta_{1}^{2}\Gamma^{2}(1+1/\beta_2)
  \end{array}
\right. \nonumber\\ & \nonumber\\ \Rightarrow&\left\{
                     \begin{array}{ll}
                         \vspace*{0.3 cm}
                       \beta_1=\frac{\mu}{\Gamma(1+1/\beta_2)}\\
                       2\log\Gamma(1+1/\beta_2)-\log\Gamma(1+2/\beta_2)+\log\frac{\mu^2+\sigma^2}{\mu^2}=0.
                     \end{array}
                   \right. &
\end{align}
\end{itemize}
In order to find $\beta_2$, the Newton-Rapson method can be used to solve the nonlinear equation. Here, we use the \verb"nleqslv" function in the \verb"R" package of the same name.

Assuming independence, the joint prior density function of $\boldsymbol{\theta}=(\mu,\sigma^2,\boldsymbol{p})$ is given by $\pi(\boldsymbol{\theta})=\pi_1(\boldsymbol{p})\pi_2(\mu)\pi_3(\sigma^2)$. Therefore, according to the Bayesian paradigm, the posterior density of $\boldsymbol{\theta}$ is
\begin{equation}\label{joint posterior}
f(\boldsymbol{\theta}|y)\propto L(y,\boldsymbol{\theta})\pi(\boldsymbol{\theta}).
\end{equation}

In this paper, the prior distributions for the connecting parameters, $\mu$ and $\sigma^2$, are assumed to be independent gamma distributions, both with a mean of one and a variance of 100, that is, $\mu, \sigma^2\sim gamma(0.01,100)$ \citep{Pereira17}. For the mixture weights, we use a Dirichlet prior, $\boldsymbol{p}\sim Dir(1,1,1)$ when all families of models are considered ($m=3$) or a Beta prior with parameters (1,1) (uniform$(0,1)$) for any combination of $m=2$.

In order to measure the evidence in favour of each model, the hypotheses on the mixture weights are tested \citep{kamary14,Pereira17}.

The hypothesis specifying that $y$ has the density function $f_k(y|\psi_k)$ is equivalent to
\begin{equation}\label{hp1}
H_k: p_k=1 \wedge p_i=0, i\neq k.
\end{equation}

On the other hand, the hypothesis that $y$ has not the density $f_k(y|\psi_k)$ is equivalent to
\begin{equation}\label{hp0}
H: p_k=0 \wedge \displaystyle\sum_{i\neq k} p_i=1.
\end{equation}

The alternative hypotheses to (\ref{hp1}) and (\ref{hp0}) are $A_k: p_k<1$ and $A_k: p_k>0$, respectively, which are not sharp anyway.

The FBST procedure is used to test $H_k,k=1,\ldots,m$, according to the expressions (\ref{tangent}) and (\ref{ev}). For the optimization step, we used the conjugate gradient method \citep{Fletcher}. In order to perform the integration over the posterior measure, we used an Adaptive Metropolis Markov chain Monte Carlo algorithm, MCMC, of \citet{Haario01}.

In this paper, the implementation of the Bayesian models is carried out using \verb"LaplacesDemon" \verb"R" package. The \verb"LaplacesDemon" is an open-source package that provides a complete environment for simulation in Bayesian inference \citep{LCC16}.

\section{Simulations}\label{sec5}
In this section we present some numerical results based on simulated sample points in order to evaluate the performance of the FBST for discriminating between separate families of distributions. Our main interest is to measure the convergence rate of correct decisions, concerning the acceptance/rejection of the true/false distribution of the generated sample, when using the FBST on the mixture model. In this paper, the simulation study is carried out in two parts. First, we compare the empirical results of the FBST and Cox test on discriminating between two separate models. Second, we apply the lognormal-gamma-Weibull mixture model (LGW) to the simulated data in order to evaluate the performance of the FBST on identifying the true distribution used to generate the sample.

The simulations of this paper were performed on a Intel(R) Core(TM) i7-5500U CPU@ 2.40GHz computer.

\subsection{Discriminating between two separate models}

\vspace*{.3 cm}
\noindent \textbf{Simulation scheme of sample points}
\vspace*{.30 cm}

In this paper, we have illustrated the simulations of the lognormal and Weibull distributions. Let $H_L$ and $H_W$ be the hypotheses specifying the probability density functions of the lognormal and Weibull models, respectively, as defined in section \ref{sec2}. For each hypothesis, we generate $500$ samples of sizes $n=20$, $40$, $60$, $80$, $100$, $150$ and $200$ from the distributions and, for every sample data, $n$, we compute the evidence in favor of the hypothesis using the FBST procedure and Cox test. Due to the invariance of the e-value \citep{Madruga} and of the maximum likelihood ratio \citep{Pereira78}, this case did not required changes in parameters values for the simulations. Therefore, the various sample sizes $n$ from lognormal were obtained with $\alpha_1=0$ and $\alpha_2=1$ ($LN(0,1)$) and the samples from Weibull were generated with $\beta_1=1$ e $\beta_2=1$ ($W(1,1)$).

As acceptance/rejection threshold, we adopted the critical level $c$ according to criterion presented in section \ref{sec3}, with a significance level of $5\%$. We chose this asymptotic criterion because of our benchmark (the Cox test) which is an asymptotic procedure as well.  Since the mixture model and the restricted model have 3 and 2 degree of freedom, respectively, we have $c=F_{3}[F_{2}^{-1}(0.95)]=0.72$. Therefore, we reject $H$ if $\overline{\mbox{ev}}(H)>0.72$ or, equivalently, if $\mbox{ev}(H)<0.28$.

For the Cox test, adopting a significance level of $5\%$, we define the rejection region as follows: $R=\{y: \big|T^{*}\big|>1.96\}$, where $T^{*}\sim N(0,1)$. The expressions for the computations of the Cox's test statistics are given in section \ref{sec2}.

\vspace*{.3 cm}
\noindent \textbf{Simulation results}
\vspace*{.30 cm}

The simulation results are summarized in the tables shown below. As expected, both the FBST and Cox test have achieved high acceptance rates of the null hypotheses that specify the true distributions used to generate the samples (see Tables \ref{AccepL_LW} and \ref{AccepW_LW}). The type I error rates (rejection rates of the true model) obtained by FBST are always below the predefined significance level ($5\%$). The significance levels attained from the Cox test are, in general, very close to $5\%$. This is what would be hoped in a specific application \citep{Pereira78}.

Regarding the rejection of the hypotheses that specify the false models, it is clear from the Tables \ref{RejW_LW} and \ref{RejL_LW} that, as the sample size increases, the rejection rate converges to $1$. The rejection rates obtained from the FBST are higher than those of the Cox test mainly when sample sizes are small. This means that the FBST presents higher discrimination power compared to the Cox test. Note that the Cox's asymptotic tests are developed under assumption that a higher power for the alternative hypothesis is required \citep{Cox61}.

\begin{table}[htp!]
\caption{Acceptance rates of true null hypothesis $H_L$. Data from $LN(0,1)$} \label{AccepL_LW}
\begin{tabular}{lccccccc} \hline
$n$&20&40&60&80&100&150&200 \\ \hline
FBST&0.992&0.978&0.984&0.980&0.984&	0.982&0.994 \\
Cox test&0.988&0.980&0.968&0.972&0.964&0.974&0.974\\ \hline
\end{tabular}
\end{table}

\begin{table}[htp!]
\caption{\small Rejection rates of false null hypothesis $H_W$. Data from $LN(0,1)$} \label{RejW_LW}
\begin{tabular}{lccccccc} \hline
$n$&20&40&60&80&100&150&200 \\ \hline
FBST&0.316&0.608&0.704&0.930&0.998&1.000&1.000\\
Cox test&0.160&0.384&0.670&0.822&0.938&1.000&1.000\\ \hline
\end{tabular}
\end{table}

\begin{table}[htp!]
\caption{Acceptance rates of true null hypothesis $H_W$. Data from $W(1,1)$} \label{AccepW_LW}
\begin{tabular}{lccccccc} \hline
$n$&20&40&60&80&100&150&200 \\ \hline
FBST&0.998&0.966&0.972&0.982&0.976&0.982&0.984 \\
Cox test&0.994&0.972&0.964&0.978&0.950&0.944&0.936\\ \hline
\end{tabular}
\end{table}

\begin{table}[htp!]
\caption{Rejection rates of false null hypothesis $H_L$. Data from $W(1,1)$} \label{RejL_LW}
\begin{tabular}{lccccccc} \hline
$n$&20&40&60&80&100&150&200 \\ \hline
FBST&0.410&0.784&0.860&0.928&0.956&0.990&1.000\\
Cox test&0.304&0.580&0.774&0.896&0.942&0.994&1.000\\ \hline
\end{tabular}
\end{table}

\subsection{Discriminating based on the LGW mixture model}

Let $H_L$, $H_G$ and $H_W$ be the hypotheses specifying the probability density functions of the lognormal, gamma and Weibull distributions, respectively. From each distribution, we generate $200$ samples of sizes $n=25$, $50$, $100$, and $200$ and, for every sample, we use the FBST on the LGW mixture model in order to compute the evidence measures in favor of the models specified in the hypotheses.

\vspace*{.3 cm}
\noindent \textbf{Criteria for evaluating the performance of the FBST}
\vspace*{.30 cm}

In order to evaluate the performance of the FBST on selecting the true distribution used to generate the sample, we have compared the measures of evidence in favor of the hypotheses $H:p_k=0$ and $H:p_k=1$, $k=L,G,W$, where $p_k$ are respectively the mixture weights associated with the lognormal, gamma and Weibull components on the LGW mixture model.

For instance, suppose that the sample has a lognormal distribution. We consider that the FBST has made a correct choice on the LGW model, if the evidence in favor of $H:p_L=0$ is less than that in favor of $H:p_G=0$ and $H:p_W =0$, and the evidence in favor of $ H:p_L= 1$ is greater than that in favor of $H:p_G=1 $ e $H:p_W=1$. The calculation of the proportions of correct decisions made by FBST is based on $200$ replicates. An analogous procedure is employed when the samples are generated from to the gamma or the Weibull distributions.

In these simulations, we have assigned $\mu=20$ and $\sigma^2=50$.

\newpage
\noindent \textbf{Simulation results}
\vspace*{.30 cm}

\begin{table}[htp!]
\caption{\small Mean of estimates for LGW model parameters and percentages of correct decisions made by FBST on selecting the true distribution of the generated samples} \label{LGW simulation}
{\small
\begin{tabular}{cccccccc} \hline
\multirow{2}{*}{Model}&\multirow{2}{*}{$n$}&$\mu$&$\sigma^2$&$p_L$&$p_G$&$p_W$&\multirow{2}{*}{$\%$ of Cd$^{*}$} \\ \cline{3-7}
                &                       &$20$&$50$&-&-&-&\\ \hline
\multirow{4}{*}{Lognormal}&$25$&$19.93$&$51.74$&$0.39$&$0.35$&$0.26$&$55$ \\
                          &$50$&$19.89$&$49.81$&$0.44$&$0.35$&$0.21$&$67$ \\
                          &$100$&$19.98$&$48.52$&$0.49$&$0.35$&$0.16$&$68$ \\
                          &$200$&$20.02$&$48.79$&$0.58$&$0.31$&$0.11$&$80$ \\ \cline{2-8}
\multirow{4}{*}{Gamma}    &$25$&$20.22$&$58.47$&$0.34$&$0.34$&$0.32$&$25$ \\
                          &$50$&$19.96$&$53.14$&$0.36$&$0.35$&$0.29$&$32$\\
                          &$100$&$20.02$&$51.80$&$0.37$&$0.38$&$0.25$&$46$\\
                          &$200$&$20.02$&$51.23$&$0.37$&$0.41$&$0.22$&$51$\\ \cline{2-8}
\multirow{4}{*}{Weibull}  &$25$&$20.26$&$59.95$&$0.28$&$0.32$&$0.40$&$71$ \\
                          &$50$&$20.09$&$54.19$&$0.24$&$0.30$&$0.46$&$80$\\
                          &$100$&$20.07$&$52.07$&$0.18$&$0.25$&$0.57$&$90$\\
                          &$200$&$20.05$&$50.89$&$0.13$&$0.20$&$0.67$&$96$ \\ \hline
                       \multicolumn{8}{l}{\small$*$ percentage of correct decision}
\end{tabular}
}
\end{table}

Table \ref{LGW simulation} presents the mean of the estimates for the LGW mixture model parameters and the percentages of correct decisions made by FBST on selecting the true distribution used to generate the samples. It is observed that, regardless of the distribution used for generating the data and the sample sizes, the estimates for the mean $\mu$ are very close to each other and to the true value of the parameter. For the estimates of the variance $\sigma^2$, we observe a variation between them but, in general, they approach the true value of the parameter as the sample size increases.

We also observed that the FBST presents good performance on identifying the Weibull distribution as the true data generation process and low performance on identifying the gamma distribution. This happens because, regarding the parameters chosen for these simulations, the gamma and lognormal densities are very similar.

\section{Applications}\label{sec6}

In this section we analyze two uncensored datasets and use the FBST and the Cox test to discriminate between lognormal, gamma and Weibull distributions.

Let us consider again the probability densities specified in the hypotheses $H_L$, $H_W$ and $H_G$. Here, the goal is to decide which of these alternative models best fits the datasets.
\newpage
\begin{Ex}
\normalfont
\citet{Kent} present a method for selecting  the member of a collection of families of distribution that best fit a set of observations. A selection statistic is proposed that is essentially the value of the density function of a scale transformation maximal invariant. The dataset observations consist of experiments for testing the tensile fatigue characteristics of polyester/viscose yarn to study the problem of warp breakage during weaving. The experiment consisted of placing $100$ samples of yarn into a $10$-station testing apparatus that subject the yarn to $80$ cycles per minute of a given strain level. The cycle at which the yarn failed (cycles-to-failure) was recorded. The FBST and the Cox test are used to compare the distributions for the data from the experiment at the 2.3 percent strain level.

Table \ref{kent} presents the Bayesian and the classical measures of evidence provided by the yarn data in favor of null hypothesis on the comparisons between pairs of the distributions. For selecting between the lognormal and the Weibull distributions, we have the following results: the e-values $ev(H_L)=0.000$ and $ev(H_W)=0.871$, and the values of standard normal deviate for Cox's test statistics $T_{LW}^{*}=-3.048$ and $T_{WL}^{*}=-0.549$ with the corresponding p-values of $0.002$ and $0.583$, respectively. These results indicate rejecting the lognormal distribution and choose the Weibull distribution which provides the best fit to the dataset. In \citet{Kent}, the Weibull distribution is also preferred over lognormal distribution. \citet{Araujo07} used the Intrinsic and Fractional Bayes factors to discriminate between these distributions and also obtained a very strong evidence against the lognormal distribution.

Since the comparison among the lognormal and gamma distributions suggests rejecting the lognormal model, the gamma versus Weibull distributions were tested. The results of the tests indicate that both the distributions provide good fit to the dataset. Again we agree with the findings of \citet{Kent} and \citet{Araujo07} which observed that it would be difficult to distinguish between those two models because both families of distributions fit these data equally well.

\begin{table}[htp!]
\caption{Measures of evidence provided by yarn data} \label{kent}
\begin{tabular}{lcccc} \hline
\multirow{3}{*}{Comparison}&\multirow{3}{*}{Null hypothesis}&\multicolumn{3}{c}{Evidence in favor of null hypothesis} \\ \cline{3-5}
                           &  &e-value& Standard normal& p-value \\
                           &  & (FBST)& deviate, $T_{fg}^{*}$&(Cox test)\\ \hline
\multirow{2}{*}{$H_L \times H_W$}&$H_L$ &$0.000$ &$-3.048$ &$0.002$ \\
                              &$H_W$ & $0.871$&$-0.549$ & $0.583$ \\ \hline
\multirow{2}{*}{$H_L \times H_G$}&$H_L$&$0.000$ & $-3.033$&$0.002$\\
                                 &$H_G$& $0.997$& $0.773$ &$0.439$\\ \hline
 \multirow{2}{*}{$H_G \times H_W$}&$H_G$& $0.697$& $1.016$&$0.309$ \\
                                  &$H_W$& $0.725$&$0.967$ & $0.333$  \\ \hline
\end{tabular}
\end{table}

In order to test simultaneously the three hypotheses, we have applied the LGW mixture model,
\begin{equation} \label{LGW1}
f(y|p,\mu,\sigma)=p_1f_L(y|\mu,\sigma)+p_2f_{G}(y| \mu, \sigma)+p_3f_W(y|\mu,\sigma),
\end{equation}
to the yarn data from the experiment at the 2.3 percent strain level.

Table \ref{LGWEst} presents the estimates for the parameters of the model (\ref{LGW1}). Here, SD, $2.5\%$ and $97.5\%$ denote the standard deviation, the $2.5$th and the $97.5$th percentiles of the posterior distribution of the LGW parameters, respectively. Table \ref{LGWTestp} gives the results of hypothesis testing on the mixture weights. The p-values are calculated according to \citet{Diniz12}, as described in section \ref{sec3}. The results of the tests are similar to the previous comparisons between pairs of the distributions. Both the classical and the Bayesian measures of evidence indicate that, among the three models, the lognormal model is the one that should not be considered because the null hypothesis $H:p_1=0$ is not rejected.

Figure \ref{Survfunctions1} displays the survival curves calculated using Bayesian estimates of the Weibull model (Table \ref{WeibullEst}), the LGW mixture model (Table \ref{LGWEst}) and a procedure called the piecewise exponential estimator (PEXE), introduced by \citet{Kim91}, representing the observed data. Unlike the well-known Kaplan-Meier estimator, the PEXE is smooth and continuous estimator of the survival function. It appears that the Weibull model by itself produces a good estimate of survival function.

\begin{table}[htp!]
\caption{Summary of the posterior distribution of the LGW parameters} \label{LGWEst}
\begin{tabular}{lcccccc} \hline
Parameter&Mean&SD&$2.5\%$&Median&$97.5\%$ \\ \hline
$p_1\mbox{-lognormal}$&$0.170$ &$0.127$&$0.007$&$0.143$&$0.469$ \\
$p_2\mbox{-gama}$&$0.381$ &$0.249$&$0.018$&$0.355$&$0.869$ \\
$p_3\mbox{-Weibull}$&$0.449$ &$0.237$&$0.032$&$0.461$&$0.869$ \\
$\mu$&$220.423$&$14.239$&$193.679$&$219.966$&$249.759$\\
$\sigma^2$&$20665.944$&$4001.274$&$14369.620$&$20248.042$&$30166.612$\\\hline
\end{tabular}
\end{table}

\begin{table}[htp!]
\caption{Hypothesis testing on the mixture weights of LGW model} \label{LGWTestp}
\begin{tabular}{lcc} \hline
Hypothesis&e-value&p-value$^*$\\\hline
$p_1=0$&$0.652$& $0.116$\\
$p_2=0$&$0.206$&$0.015$\\
$p_3=0$&$0.073$&$ 0.003$\\\hline
\multicolumn{3}{l}{\small*p-value calculated according to \citet{Diniz12}}
\end{tabular}
\end{table}

\begin{table}[htp!]
\caption{Summary of the posterior distribution of Weibull parameters} \label{WeibullEst}
\begin{tabular}{lcccccc} \hline
Parameter&Mean&SD&$2.5\%$&Median&$97.5\%$ \\\hline
$\mu$&$220.409$ &$13.675$&$194.595$&$219.938$&$248.975$\\
$\sigma^2$&$19862.278$&$3170.874$&$14708.324$&$19523.516$&$27038.00$\\ \hline
\end{tabular}
\end{table}

\begin{figure}[htp!]
\begin{center}
\includegraphics[scale=0.40]{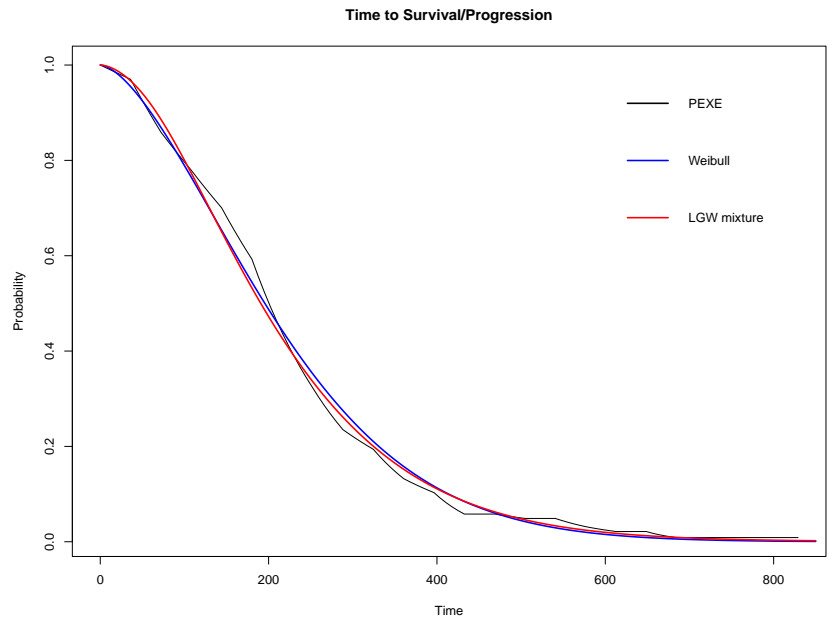}
\end{center}
\caption{Survival curves based on the estimates of the  Weibull model, the LGW model and PEXE for yarn data}  \label{Survfunctions1}
\end{figure}

The results from Tables \ref{kent} and \ref{LGWTestp} show that the preference for the Weibull model is quite clear in evaluating the 3-component mixture model more than in the 2-component model (comparison $H_G \times H_W$), where the evidence measures in favor of both models are very close. It means that the discrimination power provided by LGW model is much higher than the power of the pairwise comparisons. This finding is in agreement with the discussion of  \citet{Sawyer84}.

\end{Ex}

\begin{Ex}
\normalfont
\citep{Lagakos} This dataset are the induction times of AIDS in patients infected by contaminated blood transfusions. The times are for $258$ adults and $37$ children (less than $5$ years), infected until June $30$, $1985$, given by US Center for Disease Control.

\citet{Pereira97} analyzed the data of the adult population ($n=258$) and used the Cox tests to discriminate between $H_G\times H_W$ which have indicated that the Weibull distribution is preferable. \citet{Araujo07} used the Intrinsic and Fractional Bayes factors to discriminate between these distributions and also obtained a positive evidence against the gamma distribution.

Here, the LGW model is applied to the data of induction times for $258$ adults and the FBST is used to discriminate between the distributions by testing the hypotheses on the mixture weights. The results given by Table \ref{LGW2} indicate that neither the lognormal and gamma models should be considered because the null hypotheses $H_L:p_1=0$ and $H_G:p_2=0$ are not rejected. Consequently, among the three models, the Weibull model should be chosen for further analyses of the data. From Figure \ref{Survfunctions2} it seems reasonable to disregard both the lognormal and gamma models, since the Weibull model by itself produces a good estimate of the survival function.

\begin{table}[htp!]
\caption{Hypothesis testing on the mixture weights of LGW model} \label{LGW2}
\begin{tabular}{lcc} \hline
Hypothesis&e-value&p-value$^*$\\\hline
$p_1=0$&$0.834$& $0.227$\\
$p_2=0$&$0.856$&$0.249$\\
$p_3=0$&$0.078$&$0.004$\\\hline
\multicolumn{3}{l}{\small*p-value calculated according to \citet{Diniz12}}
\end{tabular}
\end{table}

\begin{figure}[htp!]
\belowcaptionskip = -20pt
\begin{center}
\includegraphics[scale=0.40]{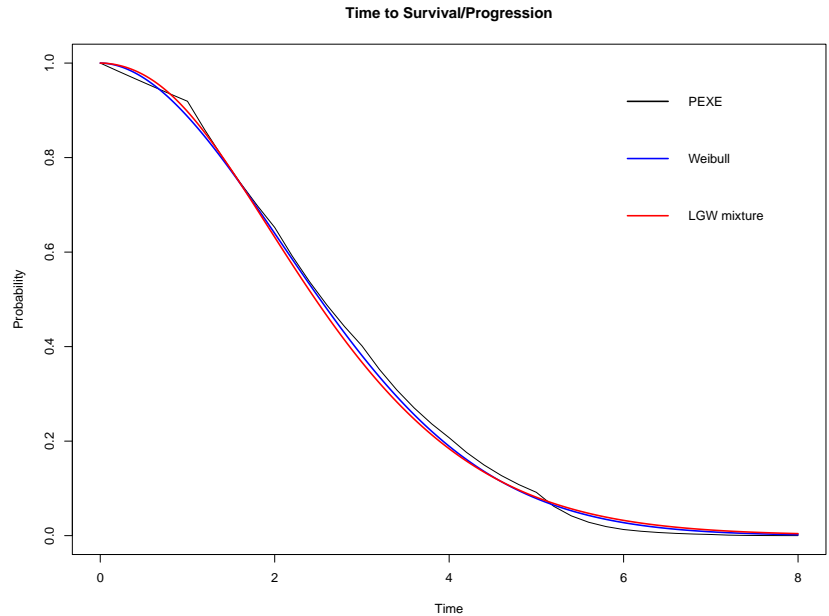}
\end{center}
\caption{Survival curves based on the estimates of the Weibull, the LGW model and the PEXE for induction times of AIDS}  \label{Survfunctions2}
\end{figure}
\end{Ex}

\newpage
\vspace*{0.0 cm}
\section{Final Remarks}\label{sec7}
In this paper we considered the FBST for discriminating between separate families of distributions. We analyzed this problem in the context of linear mixture of the candidate models. The families of distributions considered include the lognormal, gamma and Weibull models.  An advantage of using the FBST procedure for discriminating between the separate models  is that it allows for the use of improper priors \citep{PereiraSW}.

The simulation results indicated that both the FBST and Cox test have a similar behavior on discriminating between separate models. Nevertheless, the discrimination power of the FBST is slightly higher than those of the Cox test mainly for small sample sizes. For selecting based on lognormal-gamma-Weibull mixture model, the FBST achieved good performance on identifying the true distribution used to generate the data. In the examples with real datasets, the FBST reached the same conclusion as the other selection procedures used by \citet{Kent}, \citet{Araujo07} and \citet{Pereira97}. Therefore, our proposed selection procedure can be used effectively for discriminating between separate models even when the sample size is small.

When using the FBST for discriminating between separate models, it is recommended to apply a mixture model including all candidate models in order to avoid the problems that arise when pairwise comparisons are performed \citep{Sawyer84}. Whenever passible, we also recommend reparametrizing the models in terms of the common parameters.

It would be of interesting to compare the proposed procedure with other selection procedures that allow the use of data with censoring mechanisms.

\section*{Acknowledgements}
The authors are grateful for the support of CNPq, COPPE/UFRJ and IME/USP.

\end{document}